\documentclass[plb,preprint,tightenlines,floatfix,showpacs,preprintnumbers,nofootinbib]{revtex4}
  \usepackage{amssymb}
   \usepackage{amsmath}
    \usepackage{epsfig}
     \usepackage{bm}
      \usepackage{pifont}
\def\Li{\relax\ifmmode{\textbf{Li}_{2}}\else{Li$_2${ }}\fi}

\usepackage[usenames,dvipsnames]{xcolor}
\usepackage{fancybox}
\usepackage{amsmath}
\usepackage{empheq}
\usepackage{textpos}
\usepackage{pgfplots}
\usepackage{tikz}
\usepackage{dsfont}
\usepackage[nice]{nicefrac}
\usepackage{float}
\usepackage{enumerate}
\usetikzlibrary{calc}
\usetikzlibrary{shapes,arrows,decorations.markings}
\usepackage{hyperref}

\makeatletter

\AtBeginDocument{}
\AtBeginDocument{}
\AtBeginDocument{}
\newcommand{\lyxmathsym}[1]{\ifmmode\begingroup\def\b@ld{bold}
  \text{\ifx\math@version\b@ld\bfseries\fi#1}\endgroup\else#1\fi}

\newcommand{\half}{\frac{1}{2}}

\newcommand{\es}{\frac{1}{\e^2}}

\newcommand{\be}{\begin{equation}}
\newcommand{\ee}{\end{equation}}

\newcommand{\ba}{\begin{eqnarray}}
\newcommand{\ea}{\end{eqnarray}}
\newcommand{\bg}{\begin{gather}}
\newcommand{\foma}{\end{gather}}

\newcommand{\nn}{\nonumber}

\newtheorem{theorem}{Theorem}[section]
\newtheorem{lemma}[theorem]{Lemma}
\newtheorem{proposition}[theorem]{Proposition}
\newtheorem{corollary}[theorem]{Corollary}
\newtheorem{definition}[theorem]{Definition}
\newtheorem{example}[theorem]{Example}

\newcommand{\qed}{\nobreak \ifvmode \relax \else
      \ifdim\lastskip<1.5em \hskip-\lastskip
      \hskip1.5em plus0em minus0.5em \fi \nobreak
      \vrule height0.75em width0.5em depth0.25em\fi}
      
\newcommand{\bd}{\begin{definition}}
\newcommand{\ed}{\end{definition}}      
\newcommand{\bl}{\begin{lemma}}
\newcommand{\el}{\end{lemma}}   
\newcommand{\bt}{\begin{theorem}}
\newcommand{\et}{\end{theorem}}   
\newcommand{\bc}{\begin{corollary}}
\newcommand{\ec}{\end{corollary}}
\newcommand{\bex}{\begin{example}}
\newcommand{\eex}{\end{example}}   
\newcommand{\bp}{\begin{proposition}}
\newcommand{\ep}{\end{proposition}}
\newcommand{\baa}{\begin{align}}
\newcommand{\eaa}{\end{align}}

\def\e{\epsilon}

\def\<{\langle}
\def\>{\rangle}

\def\a{\alpha}

  \def\G{\Gamma}
\def\d{\delta}  
   
\def\s{\sigma}

\def\m{\mu}
\def\n{\nu}

\def\({\left(}
\def\[{\left[}
\def\){\right)}
\def\]{\right]}

\usepackage{pgfplots}
\usepackage{tikz}
\usetikzlibrary{trees}
\usetikzlibrary{decorations.pathmorphing}
\usetikzlibrary{decorations.markings}
\usetikzlibrary{intersections,shapes.arrows}
\usepackage{refstyle}
\begin{document}
\thispagestyle{empty}

\date{\today}

\title{Evolution of Light-Like Wilson Loops with a Self-Intersection in Loop Space }
\author{T.~Mertens}
\email{tom.mertens@uantwerpen.be}
\affiliation{Departement Fysica, Universiteit Antwerpen, B-2020 Antwerpen, Belgium\\}
\author{P.~Taels}
\email{pieter.taels@uantwerpen.be}
\affiliation{Departement Fysica, Universiteit Antwerpen, B-2020 Antwerpen, Belgium\\}
\vspace {10mm}
\begin{abstract}
Recently, we proposed a general evolution equation for single quadrilateral 
Wilson loops on the light-cone. In the present work, 
we study the energy evolution of a combination of two such loops that partially overlap or have a self-intersection.
We show that, for a class of geometric variations, then evolution is consistent with
our previous conjecture, and we are able to handle the intricacies associated with the self-intersections and 
overlaps. This way, 	a step forward is made towards the understanding of loop space, 
with the hope of studying more complicated structures appearing in 
phenomenological relevant objects, such as parton distributions.
\end{abstract}

\maketitle

\newpage
\section{Introduction} 
\noindent 
A reformulation of the Ambrose-Singer theorem in a gauge theory context \cite{Giles:1981ej}
states that the holonomy variables $U_\G$ of the (gauge-)connection $\mathcal{A}_\m= A_\m^a t^a$:
\be
 U_{\G_i}=\Phi (\Gamma_i)
   =
   {\cal P} \ \exp\left[ig \oint_{\Gamma_i} \ dz^\mu {\cal A}_{\mu} (z) \right] \ ,
   \label{eq : wl_def}
\ee
where the $\G_i$ are loops, and where
$t^a$ are the generators of the Lie algebra of the gauge group 
in a certain representation (here the fundamental representation of $SU(N_c)$), 
contain the same information 
as the corresponding gauge theory. 
From this holonomy, which is some $N\times N$ matrix 
in the representation of the gauge group, a gauge invariant
variable can be obtained by taking the trace\footnote{$U_\G$ 
transforms under a gauge transformation as $U_\G^g=g^{-1}_x U_\G g_x$, where $x=\G(0)$.
The cyclicity of the trace assures that $\mbox{Tr} \ U_\G$ is gauge invariant.}. 
This trace introduces, however, extra constraints on the loops: the so-called Mandelstam constraints 
\cite{Giles:1981ej,Brandt:1982gz,Makeenko:1979pb,Mandelstam:1968hz}, 
which assure that the product of traces over holonomies is again the trace of a holonomy. 
Gauge theory can then be represented in a loop space setting, 
where the observables are now built from the vacuum expectation values of products of traces of holonomies,
referred to as Wilson loop variables 
\cite{Alday:2010zy,Alday:2007hr,Brandt:1982gz,Makeenko:1980vm,Makeenko:1979pb}:
\be
 {\cal W}_n [\Gamma_1, ... \Gamma_n]
  =
\Big \langle 0 \Big| {\cal T} \frac{1}{N_c} {\rm Tr}\ \Phi (\Gamma_1)\cdot \cdot \cdot \frac{1}{N_c}{\rm Tr}\ \Phi (\Gamma_n)  \Big| 0 \Big\rangle \ .
   \label{eq : wl}
\ee
In a previous paper \cite{cherednikov2012evolution}, 
we calculated the leading order contribution of a quadrilateral
Wilson loop on the light-cone, and introduced a new differential operator:
\be\label{eq :  diffop}
	\frac{d}{d\ln{\s}} = s\frac{d}{ds} + t\frac{d}{dt},
\ee
with $s$ and $t$ the Mandelstam variables (see section 2).
This operator was then used to derive an evolution equation for this class of Wilson loops, and inspired us to 
formulate a conjecture for a general evolution equation:

\be\label{eq : conjecture}
	\lim_{\e\to 0} \m \frac{d}{d\m}\left( \frac{d}{d \ln{\s}} \mathcal{W}_1(\G)\right) = - \sum_{cusps}\G_{cusp},
\ee
where $\e$ is defined in the dimensional regularization procedure $D=4-2\e$.
It was also shown
that the evolution of the cusp and $\Pi$-shape configurations
is consistent with this conjecture (see \cite{cherednikov2012evolution} for the details).
It turns out that the operator, Eq.   (\ref{eq : conjecture}), 
is a special case of the Fr\'echet derivative \cite{munkres1991analysis,dieudonne2008foundations}, the details 
of which will be discussed elsewhere \cite{mertens_taels2013}.

In this work, we consider some symmetrical combinations of two quadrilateral Wilson loops on the light-cone,
for which we test conjecture (\ref{eq : conjecture}). 
Put in a Wilson loop variable language: we are calculating $ {\cal W}_1 [\G]$, with $\G=\G_1\G_2$,
where the product between the loops is defined in generalized loop space 
\cite{Tavares1993pw}.
Important is that these Wilson loop
configurations exhibit intricacies, associated with the self-intersection and overlap,
that usually cause problems in a loop space approach. 
A close inspection of these intricacies indicate that one needs 
to be careful in counting cusps
along the path. We show that the effective number of cusps can deviate from 
the number one would expect from a naive counting procedure. Furthermore, 
the group structure of loop space
is confirmed by our explicit calculations.

\section{Loops and parametrization}
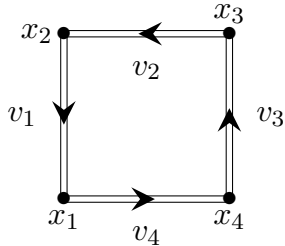
\begin{figure}[h]
	\center
	\begin{tikzpicture}[scale=1.1, transform shape]
		\tikzset{ 
		photon/.style={decorate, decoration={snake}, draw=red},    
		electron/.style={draw=blue, postaction={decorate},decoration={markings,mark=at position .5 with {\arrow[latex',draw=blue]{\arrow{latex}}}}},     
		gluon/.style={decorate, draw=magenta,     decoration={coil,amplitude=4pt, segment length=5pt}}
		};
	\tikzset{
		wilsonline/.style={draw, double distance=2pt, postaction={decorate}, decoration={markings,mark=at position .55 with 				{\arrow{stealth}}}}
		};
	\draw[wilsonline] (0,2)--node[left=.5em] {$v_1$}(0,0);
	\draw[wilsonline] (2,2)--node[below=.5em] {$v_2$}(0,2);
	\draw[wilsonline] (2,0)--node[right=.5em] {$v_3$}(2,2);
	\draw[wilsonline] (0,0)--node[below=.5em] {$v_4$}(2,0);
	
	\node (A) at (0,2) {$\bullet$};
	\node (C) at (0,0) {$\bullet$};
	\node (D) at (2,0) {$\bullet$};
	\node (G) at (2,2) {$\bullet$};

	\draw (0,0) node[anchor=north] {$x_1$};
	\draw (0,2) node[left] {$x_2$};
	\draw (2,2) node[anchor=south] {$x_3$};
	\draw (2,0) node[anchor=north] {$x_4$};
	\end{tikzpicture}
	\caption{Parametrization of the Wilson Loop in coordinate space.}
	\label{fig : parametrization}
\end{figure}
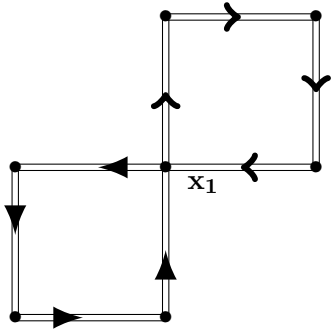
\begin{figure}[h!]
	\begin{tikzpicture}[scale=.5]
		\tikzset{
			photon/.style={decorate, decoration={snake}, draw=red},    
			electron/.style={draw=blue, postaction={decorate},decoration={markings,mark=at position .55 with 
			{\arrow[draw=blue]{>}}}},     
			gluon/.style={decorate, draw=magenta,     decoration={coil,amplitude=4pt, segment length=5pt}}};

		\tikzset{wilsonline/.style={draw, double distance=2pt, postaction={decorate}, decoration={markings,mark=at position .5 with 							{\arrow[draw=black]{to}}}}};
		\tikzset{wilsonline3/.style={draw, double distance=2pt, postaction={decorate}, decoration={markings,mark=at position .45 with 						{\arrow[draw=black]{latex}}}}};
		
		\def \sha{4}
		\def \shb{4}
		
		\draw[wilsonline] (0,0) -- (0,\shb);
		\draw[wilsonline] (0,\shb) -- (\sha,\shb);
		\draw[wilsonline]  (\sha,\shb) -- (\sha,0);
		\draw[wilsonline] (\sha,0) -- (0,0);
		\draw[wilsonline3] (-\sha,0-\shb) --  (0,0-\shb);
		\draw[wilsonline3] (-\sha,\shb-\shb) -- (-\sha,0-\shb);
		\draw[wilsonline3] (0,\shb-\shb) -- (-\sha,\shb-\shb);
		\draw[wilsonline3] (0,0-\shb) -- (0,\shb-\shb);
		\node at (0,0) {$\bullet$};
		\node[below=.6em,right=.4em] at (0,0) {$\mathbf{x_1}$};
		\node at (0,\shb) {$\bullet$};
		\node at (\sha,0) {$\bullet$};
		\node at (-\sha,0-\shb) {$\bullet$};
		\node at (-\sha,\shb-\shb) {$\bullet$};
		\node at (\sha,\shb) {$\bullet$};
		\node at (0,-\shb) {$\bullet$};
	\end{tikzpicture}
	\caption{Configuration 3}
	\label{fig : config3}
\end{figure}
\begin{figure}[h!]
	\begin{tikzpicture}[scale=.5]
		\tikzset{
			photon/.style={decorate, decoration={snake}, draw=red},    
			electron/.style={draw=blue, postaction={decorate},decoration={markings,mark=at position .55 with 
			{\arrow[draw=blue]{>}}}},     
			gluon/.style={decorate, draw=magenta,     decoration={coil,amplitude=4pt, segment length=5pt}}};

		\tikzset{wilsonline/.style={draw, double distance=2pt, postaction={decorate}, decoration={markings,mark=at position .4 with 							{\arrow[draw=black]{to}}}}};
		\tikzset{wilsonline3/.style={draw, double distance=2pt, postaction={decorate}, decoration={markings,mark=at position .45 with 						{\arrow[draw=black]{latex}}}}};
		
		\def \sha{4}
		\def \shb{4}
		
		\draw[wilsonline] (0,0) -- (0,\shb);
		\draw[wilsonline] (0,\shb) -- (\sha,\shb);
		\draw[wilsonline]  (\sha,\shb) -- (\sha,0);
		\draw[wilsonline] (\sha,0) -- (0,0);
		\draw[wilsonline3] (0,-\shb) --  (-\sha,0-\shb);
		\draw[wilsonline3] (-\sha,0-\shb) -- (-\sha,\shb-\shb);
		\draw[wilsonline3] (-\sha,\shb-\shb) -- (0,\shb-\shb);
		\draw[wilsonline3] (0,\shb-\shb) -- (0,0-\shb);
		\node at (0,0) {$\bullet$};
		\node[below=.6em,right=.4em] at (0,0) {$\mathbf{x_1}$};
		\node at (0,\shb) {$\bullet$};
		\node at (\sha,0) {$\bullet$};
		\node at (-\sha,0) {$\bullet$};
		\node at (-\sha,-\shb) {$\bullet$};
		\node at (0,-\shb) {$\bullet$};
		\node at (\sha,\shb) {$\bullet$};
	\end{tikzpicture}
	\caption{Configuration 4}
	\label{fig : config4}
\end{figure}
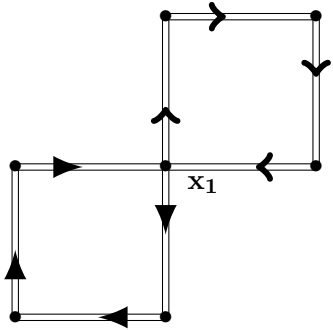

We combine two similar
quadrilateral Wilson loops on the light-cone in two different configurations: with a partial overlap
(figures \ref{fig : config1} to \ref{fig : config2}), and with a self-intersection 
(figures \ref{fig : config3} to \ref{fig : config4}).
Each loop is  parametrized by four vectors $v_i$ on the light-cone
(i.e. $v_i^2=0,\forall i$), as shown in figure \ref{fig :  parametrization}.
In the symmetric cases under consideration, both loops are equal in size and hence can 
be parametrized by these four vectors.
We also introduce the Mandelstam variables\footnote{Note the signs!}
\be
	s = (v_1 + v_2)^2 = 2v_1v_2,
	\quad\quad
	t=  (v_2 + v_3)^2 = 2v_2v_3,
\ee
in order to simplify the notation of the results of the calculation.
Each configuration has two different possible relative 
orientations for the constituting loops: one where the orientation is equal
(figures \ref{fig : config1} and \ref{fig : config4}) and one
where the orientation is opposite (figures \ref{fig : config2} and \ref{fig : config3}).
These orientations define how the Wilson line - gluon vertices are ordered along the path,
and also fix the color flow at the self-intersection or overlap.
In the figures, the color flow along the loops is represented by the different arrow styles,
and the point $\mathbf{x_1}$ represents the base-point of the considered loop space. 
Here we only consider color neutral objects, in other words: 
there are no gluons in the initial or final state.

\begin{figure}[h!]
	\begin{tikzpicture}[scale=.5]
		\tikzset{
			photon/.style={decorate, decoration={snake}, draw=red},    
			electron/.style={draw=blue, postaction={decorate},decoration={markings,mark=at position .55 with 
			{\arrow[draw=blue]{>}}}},     
			gluon/.style={decorate, draw=magenta,     decoration={coil,amplitude=4pt, segment length=5pt}}};

		\tikzset{wilsonline/.style={draw, double distance=2pt, postaction={decorate}, decoration={markings,mark=at position .4 with 							{\arrow[draw=black]{to}}}}};
		\tikzset{wilsonline3/.style={draw, double distance=2pt, postaction={decorate}, decoration={markings,mark=at position .45 with 						{\arrow[draw=black]{latex}}}}};
		
		\def \sha{4}
		\def \shb{4}
		
		\draw[wilsonline] (0,0) -- (0,\shb);
		\draw[wilsonline] (0,\shb) -- (\sha,\shb);
		\draw[wilsonline]  (\sha,\shb) -- (\sha,0);
		\draw[wilsonline] (\sha,0) -- (0,0);
		\draw[wilsonline3] (0,0) --  (-\sha,0);
		\draw[wilsonline3] (-\sha,0) -- (-\sha,\shb);
		\draw[wilsonline3] (-\sha,\shb) -- (0,\shb);
		\draw[wilsonline3] (0,\shb) -- (0,0);
		\node at (0,0) {$\bullet$};
		\node[below=.75em] at (0,0) {$\mathbf{x_1}$};
		\node at (0,\shb) {$\bullet$};
		\node at (\sha,0) {$\bullet$};
		\node at (-\sha,0) {$\bullet$};
		\node at (-\sha,\shb) {$\bullet$};
		\node at (\sha,\shb) {$\bullet$};
	\end{tikzpicture}
	\caption{Configuration 1}
	\label{fig : config1}
\end{figure}
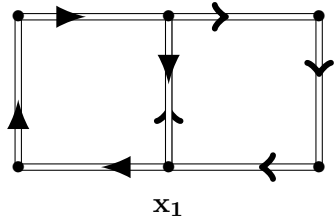
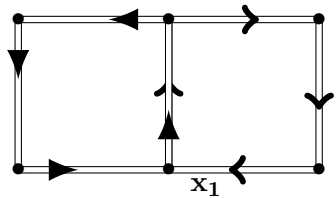
\begin{figure}[h!]
	\begin{tikzpicture}[scale=.5]
		\tikzset{
			photon/.style={decorate, decoration={snake}, draw=red},    
			electron/.style={draw=blue, postaction={decorate},decoration={markings,mark=at position .55 with 
			{\arrow[draw=blue]{>}}}},     
			gluon/.style={decorate, draw=magenta,     decoration={coil,amplitude=4pt, segment length=5pt}}};

		\tikzset{wilsonline/.style={draw, double distance=2pt, postaction={decorate}, decoration={markings,mark=at position .6 with 							{\arrow[draw=black]{to}}}}};
		\tikzset{wilsonline3/.style={draw, double distance=2pt, postaction={decorate}, decoration={markings,mark=at position .4 with 						{\arrow[draw=black]{latex}}}}};
		
		\def \sha{4}
		\def \shb{4}
		
		\draw[wilsonline] (0,0) -- (0,\shb);
		\draw[wilsonline] (0,\shb) -- (\sha,\shb);
		\draw[wilsonline]  (\sha,\shb) -- (\sha,0);
		\draw[wilsonline] (\sha,0) -- (0,0);
		\draw[wilsonline3] (-\sha,0) --  (0,0);
		\draw[wilsonline3] (-\sha,\shb) -- (-\sha,0);
		\draw[wilsonline3] (0,\shb) -- (-\sha,\shb);
		\draw[wilsonline3] (0,0) -- (0,\shb);
		\node at (0,0) {$\bullet$};
		\node[below=.6em,right=.4em] at (0,0) {$\mathbf{x_1}$};
		\node at (0,\shb) {$\bullet$};
		\node at (\sha,0) {$\bullet$};
		\node at (-\sha,0) {$\bullet$};
		\node at (-\sha,\shb) {$\bullet$};
		\node at (\sha,\shb) {$\bullet$};
	\end{tikzpicture}
	\caption{Configuration 2}
	\label{fig : config2}
\end{figure}


\section{Loop-group structure}
\noindent Expanding Eq. (\ref{eq : wl_def}) to second order
\be
	\Phi(\G) = 1 + i g \oint_{\Gamma} \ dz^\mu {\cal A}_{\mu} (z) - \frac{g^2}{2!}  {\cal P} \oint_{\Gamma} \ dz^\m dz'^\n{\cal A}_{\mu}(z) {\cal A}_{\n} (z'),
\ee
and taking traces and vacuum expectation values yields:
\be\label{eq : w1}
	\mathcal{W}_1(\G) = 1 - \frac{g^2}{2!} \mbox{Tr}(t^a t^b)
		\Big \langle 0 \Big| 
		{\cal T} \oint_{\Gamma} \ dz^\m dz'^\n A_{\mu}^a(z)  A_{\n}^b (z')
		\Big| 0 \Big\rangle \ ,
\ee
for the first order Wilson loop variable. The generators and gauge connections in Eq. (\ref{eq : w1}) are ordered along the loop by the time-ordering operation ${\cal T}$, where the ``time'' is represented by the path parameter $t\in[0,1]$ such that $dz^\m = \dot{z}^\m dt$. Now, considering $\G=\G_1\G_2$, the group structure of generalized loop space 
\cite{Tavares1993pw,chen1958,chen1968,chen1971} allows us to rewrite the integral in 
the second term of Eq. (\ref{eq : w1}) as
\be\label{eq : loop group}
	\oint\limits_{\G_1\G_2} \mathcal{A}_\m \mathcal{A}_\n = \oint\limits_{\G_1}  \mathcal{A}_\m {\cal A}_\n
	+\oint\limits_{\G_1}  {\cal A}_\m \oint\limits_{\G_2}{\cal A}_\n 
	+ \oint\limits_{\G_2}  {\cal A}_\m {\cal A}_\n \ ,
\ee
where ${\cal A}_\m$ and ${\cal A}_\n$ are again ordered along the path\footnote{This means we 
do not need to consider the contribution 
$\int\limits_{\G_1}  {\cal A}_\n \int\limits_{\G_2}{\cal A}_\m$.} and the integral measures are suppressed.
Eq. (\ref{eq : loop group}) makes it clear that there are three contributions: 
two coming from the loops considered independently, 
and one coming from the interference terms. 
Also, the group structure of generalized loop space\footnote{
The group structure was confirmed by explicit calculation of the full diagrams.}
takes care of what happens 
when one changes the orientation of one of the two loops :
\ba
	\oint\limits_{\G_1\G_2^{-1}} {\cal A}_\m {\cal A}_\n &=& \int\limits_{\G_1}  {\cal A}_\m {\cal A}_\n
	+\int\limits_{\G_1}  {\cal A}_\m \int\limits_{\G_2^{-1}}{\cal A}_\n 
	+ \int\limits_{\G_2^{-1}}  {\cal A}_\m {\cal A}_\n\nn\\
	&=&\int\limits_{\G_1}  {\cal A}_\m {\cal A}_\n
	+ (-1)^1\int\limits_{\G_1}  {\cal A}_\m \int\limits_{\G_2}{\cal A}_\n 
	+ (-1)^2\int\limits_{\G_2}  {\cal A}_\n {\cal A}_\m \ .
	\label{eq : inverseloop}
\ea 
In the last term, the order of the algebra 
generators inside the trace  needs to be reversed as well (i.e. $\mbox{Tr}(t^at^b) \to \mbox{Tr}(t^bt^a)$).
However, due to the cyclicity of the trace, at one-loop level both traces yield the same result.
Moreover, in the first order, the value of the Wilson loop variables of the single loops: the first and last term in (\ref{eq : loop group}), are independent of the orientation of the loop. Therefore, we can immediately write down the single loop contributions to Eq. (\ref{eq : inverseloop}) from our previous results in \cite{cherednikov2012evolution} (see Eq. (\ref{eq : wlsingle})).
%
\section{Diagram results}
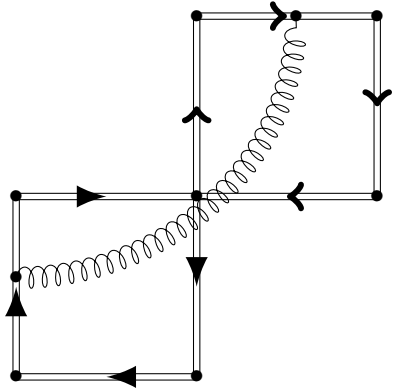
\begin{figure}[h!]
	\begin{tikzpicture}[scale=.6]
	\tikzset{
			photon/.style={decorate, decoration={snake}, draw=red},    
			electron/.style={draw=blue, postaction={decorate},decoration={markings,mark=at position .55 with 
			{\arrow[draw=blue]{>}}}},     
			gluon/.style={decorate, draw=magenta,     decoration={coil,amplitude=4pt, segment length=5pt}}
			,gluon2/.style={decorate, draw=black,     decoration={coil,amplitude=4pt, segment length=5pt}}};

		\tikzset{wilsonline/.style={draw, double distance=2pt, postaction={decorate}, decoration={markings,mark=at position .5 with 							{\arrow[draw=black]{to}}}}};
		\tikzset{wilsonline3/.style={draw, double distance=2pt, postaction={decorate}, decoration={markings,mark=at position .5 with 						{\arrow[draw=black]{latex}}}}};
		
	\def \sha{4}
		\def \shb{4}
		
		\draw[wilsonline] (0,0) -- (0,\shb);
		\draw[wilsonline] (0,\shb) -- (\sha,\shb);
		\draw[wilsonline]  (\sha,\shb) -- (\sha,0);
		\draw[wilsonline] (\sha,0) -- (0,0);
		\draw[wilsonline3] (0,-\shb) --  (-\sha,0-\shb);
		\draw[wilsonline3] (-\sha,0-\shb) -- (-\sha,\shb-\shb);
		\draw[wilsonline3] (-\sha,\shb-\shb) -- (0,\shb-\shb);
		\draw[wilsonline3] (0,\shb-\shb) -- (0,0-\shb);
		\node at (0,0) {$\bullet$};
		\node at (0,\shb) {$\bullet$};
		\node at (\sha,0) {$\bullet$};
		\node at (-\sha,0) {$\bullet$};
		\node at (-\sha,-\shb) {$\bullet$};
		\node at (0,-\shb) {$\bullet$};
		\node at (\sha,\shb) {$\bullet$};
		\draw[gluon2] (-\sha,-\shb/2+.2) to[out=0,in=-90] (\sha/2+.2,\shb);
		\node at (-\sha,-\shb/2+.2) {$\bullet$};
		\node at  (\sha/2+.2,\shb) {$\bullet$};	
		\end{tikzpicture}
		\caption{Configuration 4: example of an interference diagram }
		\label{fig : example}
\end{figure}

\noindent Using the dimensionally regularized gluon propagator in the Feynman gauge and in coordinate representation:
\be\label{eq : gluonprop}
	D_{\m\n}^{ab}(z-z')=\frac{(\m^2 \pi)^\e} {4\pi^{2}} \G[1-\e]g_{\m\n}\d^{ab}
	\frac{1}{\left(-\vert z-z'\vert^2
	-i0\right)^{1-\e}}
	 ,
\ee
we obtain for a single quadrilateral loop on the light-cone
\cite{cherednikov2012evolution,Korchemskaya:1992je,Bassetto:1993xd,Drummond:2007aua}:
{\footnotesize
\ba\label{eq : wlsingle}
	\mathcal{W}_1(\G_1) = \mathcal{W}_1(\G_2) \approx 1 -  \frac{ N_c \a_s\pi^\e} {2\pi} \G[1-\e] 
	\left( \es(-s\m^2)^\e + \es(-t\m^2)^\e  -\half \ln{\left(\frac{s}{t}\right)}^2 +\ \text{finite}\right)+\mathcal{O}(\a_s^2),\nn\\
\ea
}where we took the large $N_c$ limit \cite{Makeenko:2009dw,Makeenko:2008xr,Makeenko:2004bz,Makeenko:1979pb}
 and assumed that $\G_1$
and $\G_2$ are equal in size.
The interference terms of Eq. (\ref{eq : w1}) are computed as follows: 
let us consider, for example, one of the diagrams that contribute
to the interference term of configuration 4 (figure \ref{fig : config4}), 
which is shown in figure \ref{fig : example}.
In order to calculate this diagram, we parametrize the coordinates $z$ and $z'$ from the gluon propagator along the relevant parts of the path as follows:
\be
	z = x_1 -v_1-x v_2,\ x\in[0,1]
	\qquad
	z' = x_1 + v_2 +(1-y) v_1,\ y\in[0,1],
\ee
so that the denominator of gluon propagator becomes:
\be
	\left(-(z-z')^2\right)^{1-\e} = \left(-2v_1v_2(1+x)(2-y)\right)^{1-\e}.
\ee
The integral measure $dz\ dz'$ in this case becomes $v_1v_2\ dx\ dy=\frac{s}{2\ }dx\ dy$,
hence:
\ba
	-\frac{C_F}{N_c}\frac{g^2}{2}\int\limits_{\G_1}  A_\m(z)\ dz \int\limits_{\G_2}A_\n (z') \ dz' &=& 
		-\frac{C_F}{N_c}\frac{g^2}{2}\frac{(\m^2 \pi)^\e} {4\pi^{2}} \G[1-\e] \half \int\limits_0^1\int\limits_0^1 dx\ dy 
		\frac{(-s)}{\left(-s(1+x)(2-y)\right)^{1-\e}}\nn\\
		&\approx&  -  \frac{ N_c \a_s\pi^\e} {2\pi} \G[1-\e] \half (-s)^\e \es (2^\e-1)^2,
\ea
in the large $N_c$ limit.
Similar calculations can be done for the other diagrams and 
configurations. After summing all the contributions, this results in the following expressions for the considered configurations  
in the large $N_c$ limit (the index refers to the configuration number):
{\footnotesize
\ba
	\left[
		\left\langle 0 \left\vert
			\ \oint\limits_{\G_1\G_2} A_\m A_\n
		\right\vert 0 \right\rangle
	\right]_{1}&\approx& \frac{ N_c \a_s(\pi\m^2)^\e} {2\pi} \G[1-\e]\es\Big(
		(-s)^\e
		+(-t)^\e
		-(2^\e-1)\left[(-t)^\e +(-s)^\e \right]
		\Big)
		+ \text{finite} + \mathcal{O}(\e)
			\label{eq : resconf1}\nn\\
			\\
	\left[
		\left\langle 0 \left\vert
			\ \oint\limits_{\G_1\G_2^{-1}} A_\m A_\n
		\right\vert 0 \right\rangle
	\right]_{2}&\approx&-\frac{ N_c \a_s(\pi\m^2)^\e} {2\pi} \G[1-\e]\es\Big(
		(-s)^\e
		+(-t)^\e
		-(2^\e-1)\left((-t)^\e +(-s)^\e \right)
		\Big)
		+ \text{finite} + \mathcal{O}(\e)
			\label{eq : resconf2}\nn\\
			\\	
	\left[
		\left\langle 0 \left\vert
			\ \oint\limits_{\G_1\G_2^{-1}} A_\m A_\n
		\right\vert 0 \right\rangle
	\right]_{3}&\approx& \frac{ N_c \a_s(\pi\m^2)^\e} {2\pi} \G[1-\e]\es
		\Big(
			 (-s)^\e
			-2 (-s)^\e(2^\e-1)
			+  (-s)^\e (2^\e-1)^2
		\Big)
		+ \text{finite} + \mathcal{O}(\e)
			\label{eq : resconf3}\nn\\
			\\		
	\left[
		\left\langle 0 \left\vert
			\ \oint\limits_{\G_1\G_2} A_\m A_\n
		\right\vert 0 \right\rangle
	\right]_{4}&\approx&- \frac{ N_c \a_s(\pi\m^2)^\e} {2\pi} \G[1-\e]\es
		\Big(
			(-s)^\e
			-2 (-s)^\e(2^\e-1)
			+  (-s)^\e (2^\e-1)^2
		\Big)
		+ \text{finite} + \mathcal{O}(\e)
			\label{eq : resconf4}.,	
\nn\\		
\ea}
It should be clear from Eq. (\ref{eq : inverseloop}) that, 
at one-loop order, the sign of the interference term changes 
when reversing the orientation
of one of the loops, which is indeed confirmed by Eqs. (\ref{eq : resconf1}) to (\ref{eq : resconf4}).
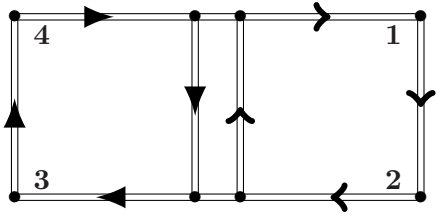
\begin{figure}[h!]
	\begin{tikzpicture}[scale =0.6]
		\tikzset{
			photon/.style={decorate, decoration={snake}, draw=red},    
			electron/.style={draw=blue, postaction={decorate},decoration={markings,mark=at position .55 with 
			{\arrow[draw=blue]{>}}}},     
			gluon/.style={decorate, draw=magenta,     decoration={coil,amplitude=4pt, segment length=5pt}}};

		\tikzset{wilsonline/.style={draw, double distance=2pt, postaction={decorate}, decoration={markings,mark=at position .5 with 							{\arrow[draw=black]{to}}}}};
		\tikzset{wilsonline3/.style={draw, double distance=2pt, postaction={decorate}, decoration={markings,mark=at position .55 with 						{\arrow[draw=black]{latex}}}}};
		
		\def \sha{4}
		\def \shb{4}
		
		\draw[wilsonline] (0,0) -- (0,\shb);
		\draw[wilsonline] (0,\shb) -- (\sha,\shb);
		\draw[wilsonline]  (\sha,\shb) -- (\sha,0);
		\draw[wilsonline] (\sha,0) -- (0,0);
		\draw[wilsonline3] (0-1,0) --  (-1-\sha,0);
		\draw[wilsonline3] (-1-\sha,0) -- (-1-\sha,\shb);
		\draw[wilsonline3] (-1-\sha,\shb) -- (-1,\shb);
		\draw[wilsonline3] (-1,\shb) -- (-1,0);
		\draw[double distance=2pt] (0,0)--(-1,0);
		\draw[double distance=2pt] (0,\shb)--(-1,\shb);
		\node at (0,0) {$\bullet$};
		\node[below=.6em,left=.3em] at (\sha,\shb) {{\color{Black}\textbf{1}}};
		\node[above=.6em,left=.3em] at (\sha,0) {{\color{Black}\textbf{2}}};
		\node[below=.6em,right=.3em] at (-1-\sha,\shb) {{\color{Black}\textbf{4}}};
		\node[above=.6em,right=.3em] at (-1-\sha,0) {{\color{Black}\textbf{3}}};
		\node at (-1,0) {$\bullet$};
		\node at (-1,\shb) {$\bullet$};
		\node at (0,\shb) {$\bullet$};
		\node at (\sha,0) {$\bullet$};
		\node at (-\sha-1,0) {$\bullet$};
		\node at (-\sha-1,\shb) {$\bullet$};
		\node at (\sha,\shb) {$\bullet$};
	\end{tikzpicture}
	\caption{Configuration 1 reduced cusps (with overlapping paths separated)}
	\label{fig : conf1 extra cusps}
\end{figure}

\begin{figure}[h!]
	\begin{tikzpicture}[scale =0.6, transform shape]
		\tikzset{
			photon/.style={decorate, decoration={snake}, draw=red},    
			electron/.style={draw=blue, postaction={decorate},decoration={markings,mark=at position .55 with 
			{\arrow[draw=blue]{>}}}},     
			gluon/.style={decorate, draw=magenta,     decoration={coil,amplitude=4pt, segment length=5pt}}};

		\tikzset{wilsonline/.style={draw, double distance=2pt, postaction={decorate}, decoration={markings,mark=at position .5 with 							{\arrow[draw=black]{to}}}}};
		\tikzset{wilsonline3/.style={draw, double distance=2pt, postaction={decorate}, decoration={markings,mark=at position .55 with 						{\arrow[draw=black]{latex}}}}};
		
		\def \sha{4}
		\def \shb{4}
		
		\draw[wilsonline] (0,0) -- (0,\shb);
		\draw[wilsonline] (0,\shb) -- (\sha,\shb);
		\draw[wilsonline]  (\sha,\shb) -- (\sha,0);
		\draw[wilsonline] (\sha,0) -- (0,0);
		\draw[double distance=2pt] (0,0)--(-1,0);
		\draw[double distance=2pt] (0,\shb)--(-1,\shb);
		\draw[wilsonline3] (-\sha-1,0) --  (0-1,0);
		\draw[wilsonline3] (-\sha-1,\shb) -- (-\sha-1,0);
		\draw[wilsonline3] (0-1,\shb) -- (-1-\sha,\shb);
		\draw[wilsonline3] (0-1,0) -- (0-1,\shb);
		\node at (0,0) {$\bullet$};
		\node[above=.6em,right=.3em] at (0,0) {{\color{Black}\textbf{1}}};
		\node[below=.6em,right=.3em] at (0,\shb) {{\color{Black}\textbf{2}}};
		\node[below=.6em,left=.3em] at (\sha,\shb) {{\color{Black}\textbf{3}}};
		\node[above=.65em,left=.3em] at (\sha,0) {{\color{Black}\textbf{4}}};
		\node[above=.65em,left=.2em] at (0,0) {{\color{Black}\textbf{5}}};
		\node[below=.6em,left=.2em] at (0,\shb) {{\color{Black}\textbf{6}}};
		\node[above=.65em,right=.2em] at (-1,0) {{\color{Black}\textbf{7}}};
		\node[below=.6em,right=.2em] at (-1,\shb) {{\color{Black}\textbf{8}}};
		\node[above=.65em,left=.3em] at (-1,0) {{\color{Black}\textbf{9}}};
		\node[below=.6em,left=.3em] at (-1,\shb) {{\color{Black}\textbf{10}}};
		\node[below=.6em,right=.3em] at (-1-\sha,\shb) {{\color{Black}\textbf{11}}};
		\node[above=.65em,right=.3em] at (-1-\sha,0) {{\color{Black}\textbf{12}}};
		\node at (-1,0) {$\bullet$};
		\node at (-1,\shb) {$\bullet$};
		\node at (0,\shb) {$\bullet$};
		\node at (\sha,0) {$\bullet$};
		\node at (-\sha-1,0) {$\bullet$};
		\node at (-\sha-1,\shb) {$\bullet$};
		\node at (\sha,\shb) {$\bullet$};
	\end{tikzpicture}
	\caption{Configuration 2 extra cusps  (with overlapping paths separated)}
	\label{fig : conf2 extra cusps}
\end{figure}
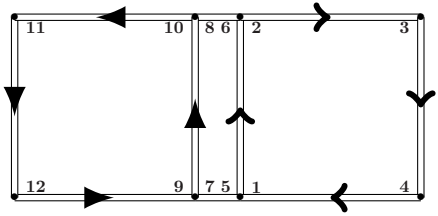 

\begin{figure}[h!]
	\begin{tikzpicture}[scale=.6]
		\tikzset{
			photon/.style={decorate, decoration={snake}, draw=red},    
			electron/.style={draw=blue, postaction={decorate},decoration={markings,mark=at position .55 with 
			{\arrow[draw=blue]{>}}}},     
			gluon/.style={decorate, draw=magenta,     decoration={coil,amplitude=4pt, segment length=5pt}}};

		\tikzset{wilsonline/.style={draw, double distance=2pt, postaction={decorate}, decoration={markings,mark=at position .5 with 							{\arrow[draw=black]{to}}}}};
		\tikzset{wilsonline3/.style={draw, double distance=2pt, postaction={decorate}, decoration={markings,mark=at position .55 with 						{\arrow[draw=black]{latex}}}}};
		
		\def \sha{4}
		\def \shb{4}
		
		\draw[wilsonline] (0,0) -- (0,\shb);
		\draw[wilsonline] (0,\shb) -- (\sha,\shb);
		\draw[wilsonline]  (\sha,\shb) -- (\sha,0);
		\draw[wilsonline] (\sha,0) -- (0,0);
		\draw[wilsonline3] (-\sha,0-\shb) --  (0,0-\shb);
		\draw[wilsonline3] (-\sha,\shb-\shb) -- (-\sha,0-\shb);
		\draw[wilsonline3] (0,\shb-\shb) -- (-\sha,\shb-\shb);
		\draw[wilsonline3] (0,0-\shb) -- (0,\shb-\shb);
		\node at (0,0) {$\bullet$};
		\node at (0,\shb) {$\bullet$};
		\node at (\sha,0) {$\bullet$};
		\node at (-\sha,0-\shb) {$\bullet$};
		\node at (-\sha,\shb-\shb) {$\bullet$};
		\node at (\sha,\shb) {$\bullet$};
		\node at (0,-\shb) {$\bullet$};
		\node[below=.6em,right=.3em] at (0,\shb) {{\color{Black}\textbf{1}}};
		\node[below=.6em,left=.3em] at (\sha,\shb) {{\color{Black}\textbf{2}}};
		\node[above=.6em,left=.3em] at (\sha,0) {{\color{Black}\textbf{3}}};
		\node[below=.6em,right=.2em] at (-\sha,0) {{\color{Black}\textbf{4}}};
		\node[above=.6em,right=.2em] at (-\sha,-\shb) {{\color{Black}\textbf{5}}};
		\node[above=.6em,left=.2em] at (0,-\shb) {{\color{Black}\textbf{6}}};

		\end{tikzpicture}
	\caption{Configuration 3 reduced number of cusps}
	\label{fig : config3 extra cusps}
\end{figure}
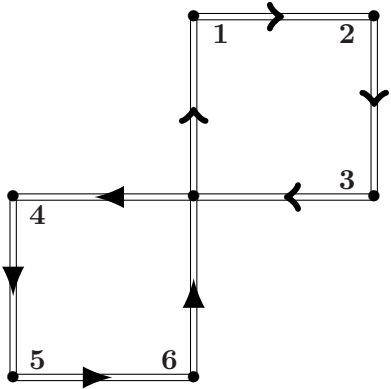

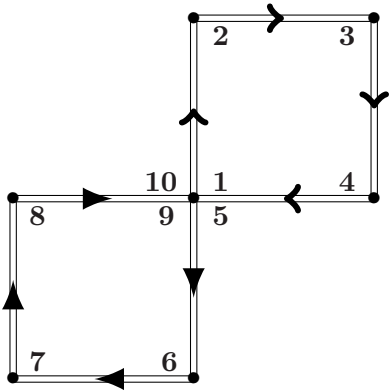
\begin{figure}[h!]
	\begin{tikzpicture}[scale=.6]
		\tikzset{
			photon/.style={decorate, decoration={snake}, draw=red},    
			electron/.style={draw=blue, postaction={decorate},decoration={markings,mark=at position .55 with 
			{\arrow[draw=blue]{>}}}},     
			gluon/.style={decorate, draw=magenta,     decoration={coil,amplitude=4pt, segment length=5pt}}};

		\tikzset{wilsonline/.style={draw, double distance=2pt, postaction={decorate}, decoration={markings,mark=at position .5 with 							{\arrow[draw=black]{to}}}}};
		\tikzset{wilsonline3/.style={draw, double distance=2pt, postaction={decorate}, decoration={markings,mark=at position .55 with 						{\arrow[draw=black]{latex}}}}};
		
		\def \sha{4}
		\def \shb{4}
		
		\draw[wilsonline] (0,0) -- (0,\shb);
		\draw[wilsonline] (0,\shb) -- (\sha,\shb);
		\draw[wilsonline]  (\sha,\shb) -- (\sha,0);
		\draw[wilsonline] (\sha,0) -- (0,0);
		\draw[wilsonline3] (0,-\shb) --  (-\sha,0-\shb);
		\draw[wilsonline3] (-\sha,0-\shb) -- (-\sha,\shb-\shb);
		\draw[wilsonline3] (-\sha,\shb-\shb) -- (0,\shb-\shb);
		\draw[wilsonline3] (0,\shb-\shb) -- (0,0-\shb);
		\node at (0,0) {$\bullet$};
		\node at (0,\shb) {$\bullet$};
		\node at (\sha,0) {$\bullet$};
		\node at (-\sha,0) {$\bullet$};
		\node at (-\sha,-\shb) {$\bullet$};
		\node at (0,-\shb) {$\bullet$};
		\node at (\sha,\shb) {$\bullet$};
		\node[above=.6em,right=.3em] at (0,0) {{\color{Black}\textbf{1}}};
		\node[below=.6em,right=.3em] at (0,\shb) {{\color{Black}\textbf{2}}};
		\node[below=.6em,left=.3em] at (\sha,\shb) {{\color{Black}\textbf{3}}};
		\node[above=.6em,left=.3em] at (\sha,0) {{\color{Black}\textbf{4}}};
		\node[above=.6em,left=.2em] at (0,0) {{\color{Black}\textbf{10}}};
		\node[below=.6em,right=.2em] at (-\sha,0) {{\color{Black}\textbf{8}}};
		\node[above=.6em,right=.2em] at (-\sha,-\shb) {{\color{Black}\textbf{7}}};
		\node[above=.6em,left=.2em] at (0,-\shb) {{\color{Black}\textbf{6}}};
		\node[below=.6em,left=.3em] at (0,0) {{\color{Black}\textbf{9}}};
		\node[below=.6em,right=.3em] at (0,0) {{\color{Black}\textbf{5}}};

		\end{tikzpicture}
	\caption{Configuration 4 extra cusps}
	\label{fig : config4 extra cusps}
\end{figure}
\section{Differential operator}
\noindent Finally, we want to apply the differential operator (\ref{eq :  diffop})
followed by the operator $\m\frac{d}{d\m}$ ,
to the considered loop configurations, after which we take the $\e\to 0$ limit.
Only symmetric variations\footnote{
Due to the fact that $s\frac{d}{ds}=(2s)\frac{d}{d(2s)}$ and 
$t\frac{d}{dt}=(2t)\frac{d}{d(2t)}$ we only need to consider $s\frac{d}{ds} + t\frac{d}{dt}$}\footnote{
Non-symmetric variations are now being studied by us, 
the results will be reported separately \cite{mertens_taels2013nonsymmvar}.} are taken into account,
in such a way that the complete loop does not changes its general structure. Indeed, if one 
does consider structure changing variations, the number of cusps change 
and in \cite{mertens2013failareader}\footnote{In preparation for publication.}
we will show explicitly that there is no continuous deformation
in loop space between two structurally different loops (i.e. loops with
a different number of cusps). 
The variations we apply here are
shown in  figures \ref{fig : deltat1} to \ref{fig : deltas2}. 
Applying the said combination of  operators to the different configurations
in the $\e\to 0$ and large $N_c$ limits yields the following results:
\ba
	\lim_{\e\to 0} \m\frac{d}{d\m} \frac{d \ln{\left(\mathcal{W}_1(\G)\right)_1}}{d\ln{\s}}
		&=& -4 \G_{cusp}\\
	\lim_{\e\to 0} \m\frac{d}{d\m} \frac{d \ln{\left(\mathcal{W}_1(\G)\right)_2}}{d\ln{\s}}
		&=& -12 \G_{cusp}\\	
	\lim_{\e\to 0} \m\frac{d}{d\m} \frac{d \ln{\left(\mathcal{W}_1(\G)\right)_3}}{d\ln{\s}}
		&=& -6 \G_{cusp}\\
	\lim_{\e\to 0} \m\frac{d}{d\m} \frac{d \ln{\left(\mathcal{W}_1(\G)\right)_4}}{d\ln{\s}}
		&=& -10 \G_{cusp},
\ea
where 
\be
	\G_{cusp} \approx \frac{\a_s N_c}{2\pi}+\mathcal{O}(\a_s^2).
\ee
Naively counting the number of cusps in the different configurations 
yields a total of eight, in each configuration. 
If this is true, however, the results contradict our conjecture 
Eq. (\ref{eq : conjecture}) \cite{cherednikov2012evolution}:
indeed,  we would expect for all the configurations a value of $-8 \G_{cusp}$ in the r.h.s.
To understand this apparent contradiction, one has to take a closer look at how to
count the number of cusps effectively present in the studied loop.
Due to path reduction \cite{Tavares1993pw,chen1958,chen1968,chen1971},
configuration 1 (figure \ref{fig : config1})\footnote{
We separated the overlapping paths in the figure to clarify the counting procedure.}
can be reduced to a single quadrilateral
so that there are effectively only four cusps (see also figure \ref{fig : conf1 extra cusps}). 
Figure \ref{fig : conf2 extra cusps} shows how count the number of cusps 
in configuration 2 (figure \ref{fig : config2}).
Here the extra cusps stem from the fact that the middle line 
is crossed twice in the same direction by the color flow 
along the path, so that these four cusps\footnote{
Four because the cusps left and right of the middle path show up in the calculations of the interference contribution.} 
contribute twice to the total number of cusps. 
Similar reasoning can be applied to configurations 3 (figure \ref{fig : config3}) and 4 (figure \ref{fig : config4}), 
where the counting is demonstrated in figures \ref{fig : config3 extra cusps} and \ref{fig : config4 extra cusps}
respectively.
\begin{figure}[h!]
	\begin{tikzpicture}[scale = .5]
		\draw (0,0) rectangle (4,4) ;
		\node at (2,4.5) {$\d t$};
		\draw (-4,0) rectangle (0,4);
		\node at (-2,4.5) {$\d t$};
		\filldraw[thick, draw=black,fill =Gray,opacity=0.4] (0,4) rectangle (4,5);
		\filldraw[thick, draw=black,fill =Gray,opacity=0.4] (-4,4) rectangle (0,5);
	\end{tikzpicture}
	\caption{$\d t$ variation of the first two configurations}
	\label{fig : deltat1}
\end{figure}
\begin{figure}[h!]
	\begin{tikzpicture}[scale = .5]
		\draw (0,0) rectangle (4,4) ;
		\node at (4.5,2) {$\d s$};
		\draw (-4,0) rectangle (0,4);
		\node at (-4.5,2) {$\d s$};
		\filldraw[thick, draw=black,fill =Gray,opacity=0.4] (4,0) rectangle (5,4);
		\filldraw[thick, draw=black,fill =Gray,opacity=0.4] (-5,0) rectangle (-4,4);
	\end{tikzpicture}
	\caption{$\d s$ variation of the first two configurations}
	\label{fig : deltas1}
\end{figure}
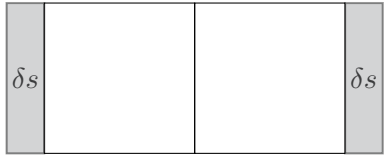
\begin{figure}[h!]
	\begin{tikzpicture}[scale = .5]
		\def \sha{4};
		\draw (0,0) rectangle (4,4) ;
		\node at (2,4.5) {$\d t$};
		\draw (-4,0-\sha) rectangle (0,4-\sha);
		\node at (-2,-4.5) {$\d t$};
		\filldraw[thick, draw=black,fill =Gray,opacity=0.4] (0,4) rectangle (4,5);
		\filldraw[thick, draw=black,fill =Gray,opacity=0.4] (-4,-\sha-1) rectangle (0,-\sha);
	\end{tikzpicture}
	\caption{$\d t$ variation of the second two configurations}
	\label{fig : deltat2}
\end{figure}
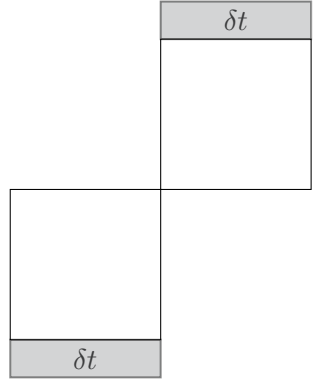
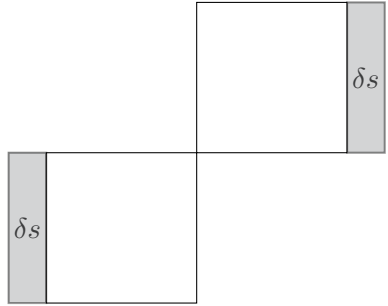
\begin{figure}[h!]
	\begin{tikzpicture}[scale = .5]
		\def \sha{4};
		\draw (0,0) rectangle (4,4) ;
		\node at (4.5,2) {$\d s$};
		\draw (-4,0-\sha) rectangle (0,4-\sha);
		\node at (-4.5,-2) {$\d s$};
		\filldraw[thick, draw=black,fill =Gray,opacity=0.4] (4,0) rectangle (5,4);
		\filldraw[thick, draw=black,fill =Gray,opacity=0.4] (-5,-4) rectangle (-4,0);
	\end{tikzpicture}
	\caption{$\d s$ variation of the second two configurations}
	\label{fig : deltas2}
\end{figure}
\section{Conclusions and remarks}
\noindent By explicit calculation, we have showed that the symmetric double Wilson loops obey the
generalized loop space group structure with respect to the inverse and the product of two loops. 
We also showed that the presented results are consistent with our evolution conjecture 
in \cite{cherednikov2012evolution}.
At the same time, we successfully dealt with some of the intricacies, 
caused by a self intersection and a partial overlap, 
that appear in calculations of Wilson loops as functionals in loop space. 
This opens the door to the calculation of more complicated structures, such as the ones appearing
in the calculations of transverse dependent momentum distributions (TMDs) or soft factors in 
certain factorization schemes in QCD \cite{Bomhof:2007xt,Cherednikov:2013bxa}. 

We would also like to point out that, when considering planar Wilson loops on the light-cone,
the number of cusps seem to behave like a winding number of some sort, in the sense that
in some cases the cusps add, and in other cases they subtract. Also note that from a loop
space point of view, loops on the light-cone with a different number of cusps can not be deformed continuously
into each other \cite{mertens2013failareader}, inducing a kind of connected component structure in loop space.
These considerations suggests a possible connection with the logarithm in the complex plane, where 
the number of times that one winds around the origin could be interpreted as a representation of the
 solution branch on which one is located. In our case, this winding would be represented by the number of
cusps, while the different connected components represent the solution branches. 
This ``duality'' will be reported separately.

\section{Acknowledgements}
We would like to thank Igor Cherednikov and Frederik Van der Veken for useful discussions and suggestions.

\bibliographystyle{h-physrev6}

\bibliography{bookreftom}	
\end{document}